# Quantum Entanglement of High Angular Momenta

Robert Fickler[1,2,*], Radek Lapkiewicz[1,2], William N. Plick[1,2], Mario Krenn[1,2], Christoph Schaeff[1,2], Sven Ramelow[1,2], Anton Zeilinger[1,2,3,*]

[1]*Quantum Optics, Quantum Nanophysics, Quantum Information, University of Vienna, Boltzmanngasse 5, Vienna A-1090, Austria*

[2]*Institute for Quantum Optics and Quantum Information, Austrian Academy of Science, Boltzmanngasse 3, Vienna A-1090, Austria*

[3]*Vienna Center for Quantum Science and Technology, Faculty of Physics, University of Vienna, Boltzmanngasse 5, Vienna A-1090, Austria*

*Correspondence to: robert.fickler@univie.ac.at or anton.zeilinger@univie.ac.at

**Abstract**:

Single photons with helical phase structures may carry a quantized amount of orbital angular momentum (OAM) and their entanglement is important for quantum information science and fundamental tests of quantum theory. Because there is no theoretical upper limit on how many quanta of OAM a single photon can carry, it is possible to create entanglement between two particles with an arbitrary high difference in the quantum number. By transferring polarization entanglement to OAM with an interferometric scheme, we generate and verify entanglement up to 600 quanta difference in the orbital angular momentum. The only restrictive factors towards higher numbers are current technical limitations. We also experimentally demonstrate that the entanglement of very high OAM can improve the sensitivity of the angular resolution in remote sensing.

Quantum entanglement - the non-classical phenomenon of joint measurements of at least two separate systems showing stronger correlations than classically explainable *(1,2)* - is widely considered as one of the most quintessential features of quantum theory. Since its discovery and first experimental demonstration *(3)* photon entanglement has been shown in various different degrees-of-freedom (DOF) *(4-7)*. A vibrant field of photonic quantum optics studies the orbital angular momentum of light (OAM). The natural solutions of the paraxial wave equation in cylindrical coordinates, Laguerre-Gauss modes, have a helical phase dependence which leads to a vortex or phase singularity and thus zero intensity along the beam axis and OAM that can take any integer value *(8)*. Entanglement of OAM of photons *(5)* led to many novel insights and applications in quantum foundations and quantum information, for example qutrit quantum communication protocols *(9)*, uncertainty relations with angular position and OAM *(10)* or higher dimensional entanglement *(11-13)*.

An important motivation for our work is the open question about the existence of macroscopic entanglement, which is of course intricately connected to the very definition of "macroscopicity" *(14)*. The OAM degree-of-freedom offers the possibility to create entanglement of quantum numbers in principle up to arbitrarily high values. In optomechanical experiments, which currently already employ linear momentum *(15)*, such photons can be used to create entanglement of mechanical system with high angular momentum. Also, quantum remote sensing offers an improved angular resolution compared to classical methods which is amplified by large OAM values *(16)*.

Due to the rapidly decreasing efficiency of the down-conversion process for the direct generation of entanglement of higher OAM *(17,18)*, we use a different approach similar to the ideas in *(7,19-21)*: We start with high-fidelity two-dimensional entanglement in the polarization DOF and transfer it to various, chosen OAM subspaces (Fig. 1). The polarization-entangled photon pairs propagating in single mode fibers enter the transfer setups in the two-photon state described by the tensor product of two degrees-of-freedom (polarization and OAM):

$$|\psi_{in}\rangle = (\alpha|H\rangle|V\rangle + e^{i\varphi}\beta|V\rangle|H\rangle) \otimes |0\rangle|0\rangle \qquad (1)$$

Where $\alpha$, $\beta$ and $\varphi$ are real and normalized $\alpha^2+\beta^2=1$, $H$ ($V$) corresponds to horizontal (vertical) polarization, 0 indicates the amount of OAM per photon (Gauss mode) and the positions of the ket-vectors label the different photons. The transfer is realized with a folded interferometric structure, which is intrinsically phase stable and has by design equal arm lengths. Depending on their polarization the photons are transferred to a well-defined LG mode by a spatial light modulator (SLM), which modulates the phase of the light *(22)*. The SLM is programmed such that, photons that take the path for H (V) polarized light are changed to $+l$ ($-l$) after leaving the interferometer.

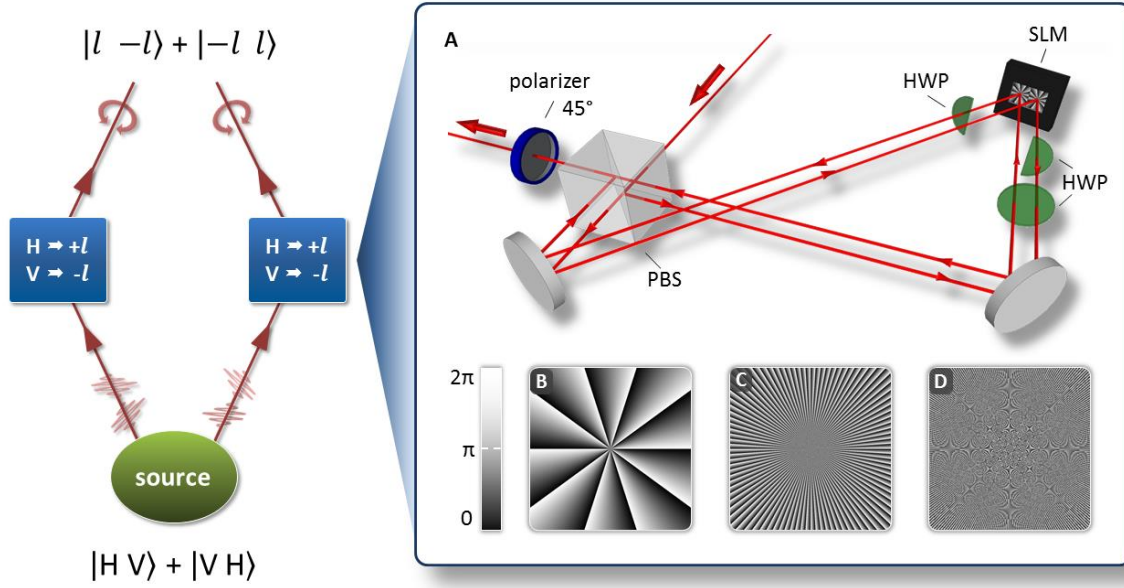

**Fig. 1**. Schematic sketch of the setup (left). Polarization entanglement is created in a parametric down-conversion process (source – oval green box) and afterwards transferred to modes with high quanta of OAM (transfer setup – rectangular blue boxes). The inset (A) shows the experimental layout for one of the two identical transfer setups where the photon is split by polarization (PBS) and its spatial mode is transformed to a higher order Laguerre-Gauss mode by a spatial light modulator (SLM). Half-wave plates (HWP) in the paths ensure that the SLM works optimally and the output is separated from the input. A polarizer (blue) projects the photon to diagonal polarization and completes the transfer. Three phase pattern, $l=10$ (B), $l=100$ (C) and $l=300$ (D), visualize the increasing complexity of the structure and the limitations because of emergence of a Moiré pattern due to the finite resolution of the SLM.

Finally, a polarizer projects the photons onto the diagonal basis (D) erasing any information about path and creating the state

$$|\psi_{out}\rangle = |D\rangle|D\rangle \otimes (\alpha|+l\rangle|-l\rangle + e^{i\varphi}\beta|-l\rangle|+l\rangle) \quad (2)$$

Our method does not rely on phase matching as in OAM-entanglement generation directly in spontaneous parametric down conversion (SPDC) but rather benefits from the efficient and high quality polarisation entanglement. The resulting states can be entangled in any two-dimensional subspace of the infinite dimensional OAM-space or even in the general space of all spatial modes which can be generated with SLMs *(23-25)*.

To realize projective measurements onto high OAM super-position states in order to demonstrate entanglement we employ a novel technique which takes advantage of the specific spatial structure of the created photonic states (Fig. 2). Any equal Laguerre-Gauss superposition, $|\chi\rangle = \frac{1}{\sqrt{2}}\left(|l\rangle + e^{i\varphi}|-l\rangle\right)$, has an intensity distribution showing a radially symmetric structure with *2l* maxima arranged in a ring (Fig. 2B-D). Because the angular position of this petal structure $\gamma = \frac{\varphi}{2l}\frac{360°}{2\pi}$ is determined by the phase $\varphi$, a mask with 2l slits maximally transmits only a specific superposition. By rotating the mask, all possible superpositions with equal amplitude are accessible. For analyzing the created state and demonstrating entanglement one mask is installed after each transfer setup and the coincidence of two transmitted photons for different combinations of $\gamma$ is measured.

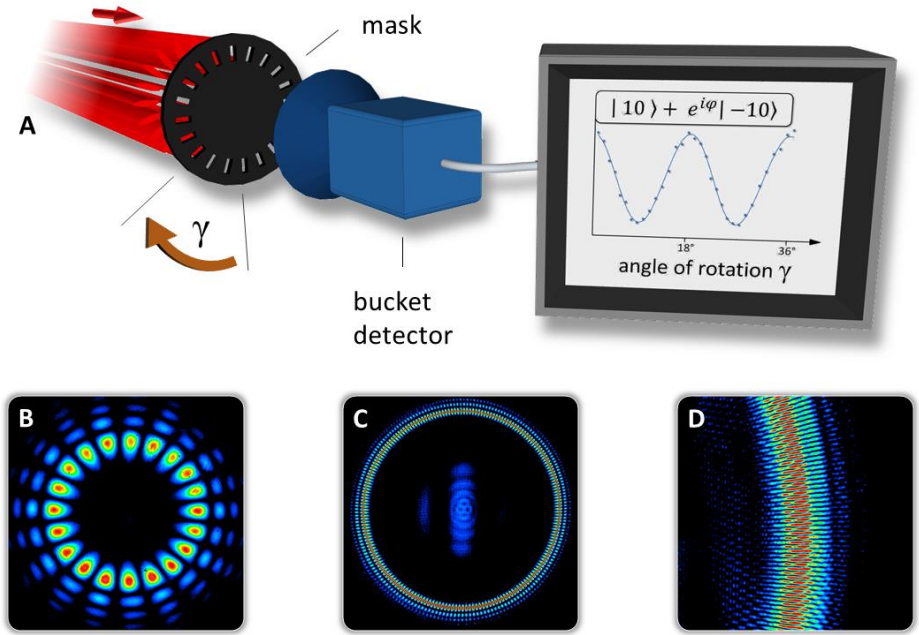

**Fig. 2.** Sketch of the measurement principle. (A) The angular position of the radial superposition structure (red) is dependent on the phase $\varphi$ of the state, here $|\chi\rangle=|10\rangle+\exp(i\varphi)|10\rangle$. A mask with the same rotational symmetry (20 slits) is able to detect any superposition dependent on its angular position $\gamma$. (B-D) show three superpositions (false colors encode intensity) for *l=±10* (B), *l=±100* (C) and a section of the mode for *l=±300* (D), that were experimentally created with a laser imaged by a CCD camera. Additional structures around and inside of the main intensity pattern (higher order LG modes with the same OAM and unmodulated photons respectively) arise from the imperfect creation of the modes at the SLM but will be blocked by the slit mask. For *l=±300,* only the shown section of the mode was used in the measurements, because of the noise from diffraction at the SLM housing and distorted modulation due to the limitations from the finite resolution.

In our experiment, polarization-entangled photon pairs (uncorrected average visibility 97.99 (3) %) at 810nm were created using a type-II nonlinear crystal in a Sagnac-type configuration *(26,27)*. The SLM in the transfer setup is programmed such that the reflected photons acquire $l$ multiples of $2\pi$ azimuthal phase ($l$ quanta of OAM) which leads to complex patterns for high OAMs (Fig. 1B-D). Therefore, we used a high resolution SLM (1920x1080 – fullHD, holoeye) with small pixel size (8µm). Nevertheless, for values of $l \geq 300$ we observed a clear reduction of mode transformation efficiency, which is the main limiting factor *(22)*. Importantly, this is only a technical limitation which can be overcome by higher resolution SLMs or novel techniques for creating photons with higher $l$ values *(28)*. After the transfer setups the modes are enlarged to fit the masks (laser-cut black cardboard) and transmitted photons are focused to bucket detectors.

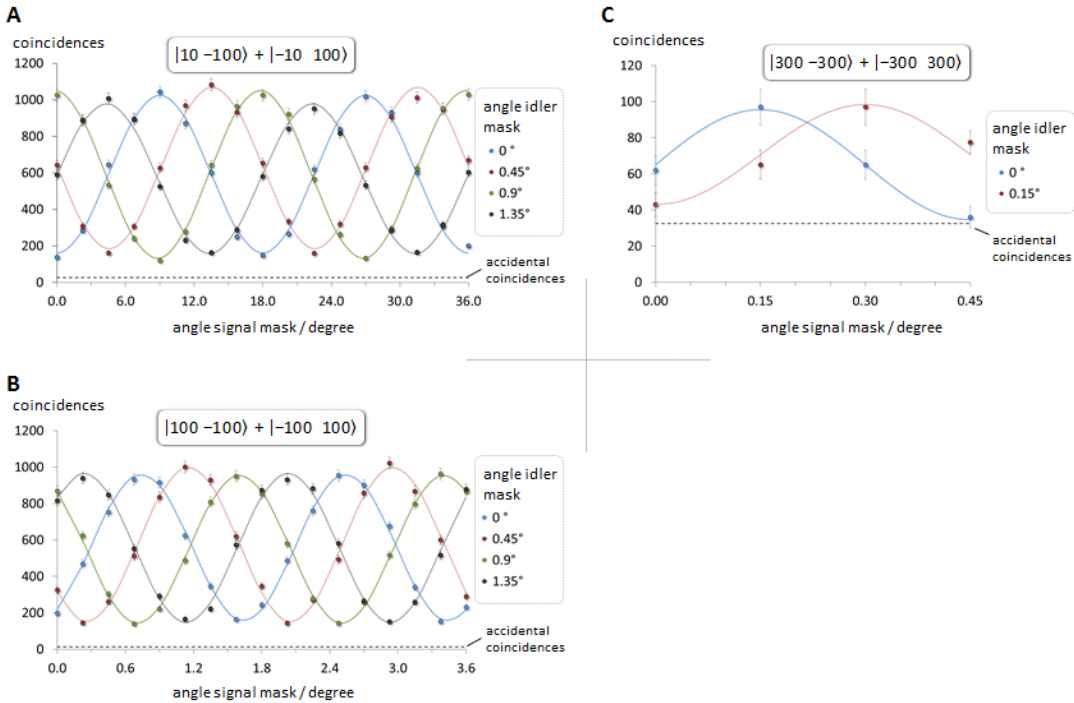

**Fig. 3.** Measured coincidence counts as a function of the angle of one mask and different angles of the other mask (different integration times for A, B and C). The measured coincidence counts (points) show a sinusoidal dependence (fitted lines) and depend only on the difference between the angles of the masks which is a clear signature of non-classical correlations. In (A) we transferred the first photon to $l=\pm 10\hbar$ and the second to $l=\pm 100\hbar$, showing the ability of creating asymmetric OAM entangled states. In (B) both photons are transferred to $l=\pm 100\hbar$. In (C) both photons carry $l=\pm 300\hbar$ and still non-classical correlations can be measured. Here, the count rate decreased significantly (approx. 1 coincidence count per minute) mainly due to limited conversion efficiency. Error bars in all plots (if big enough to be seen) are estimated from Poissonian count statistics.

First, we show the flexibility of our setup by creating two-dimensional spatial mode entanglement with highly asymmetric OAM states, where one photon is transferred to $l=\pm 10$ and the second to $l=\pm100$ (Fig. 3A). Due to its intrinsic angular momentum conservation the SPDC process could not create this asymmetric state directly. Secondly, we transfer both photons to $l=\pm100$, showing the ability to create OAM modes with very high difference in quantum number (Fig. 3B). The highest value of OAM per single photon where strong correlations were still measurable was $l=\pm300$ for both photons (Fig. 3C). The decrease in mode transformation efficiency of the SLM however strongly affects the coincidence rate (approx. 1 coincidence count per minute in the maximum) and therefore the statistical significance.

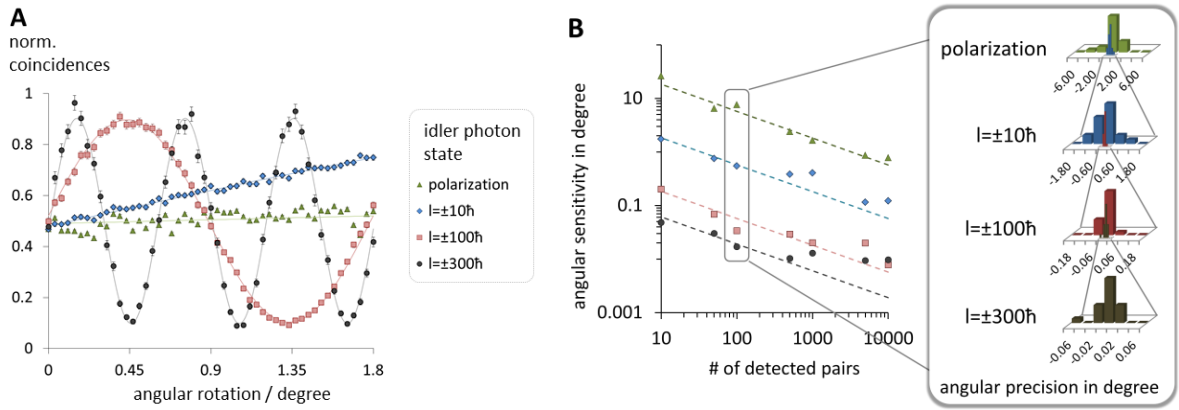

**Fig. 4.** Measurements of remote angular sensitivity enhancement. (A) Normalized coincidence count rates where the one photon is projected on diagonal polarization. The second photon is either kept polarization encoded while the polarizer is rotated (green triangles) or transferred to $l=\pm10\hbar$ (blue diamonds), to $l=\pm100\hbar$ (red squares) or to $l=\pm300\hbar$ (black circles), while the appropriate mask is rotated. The errors are estimated assuming Poissonian count statistics. From the steepest part of the fringes (0°) it is possible to calculate the corresponding angular sensitivity limited by statistical fluctuations for different numbers of detected pairs (B). The dashed lines are the theoretically expected sensitivities (assuming 100% visibility and Poissionian fluctuation) and the points are the measured values. To illustrate the enhancement for 100 detected pairs we measured the angular position of the randomly rotated mask by correcting the change in the coincidence counts with a rotation of the remote polarizer. The inset shows histograms of 20 different random angles that were measured for each arrangement. For $l=\pm300\hbar$, the limit of our high-precision rotation stage (±0.016°) was determined with the polarizer in a low-precision mount (±1°). To reach the same precision without OAM-induced angular resolution enhancement, approx. 3.3 million detected pairs would have been necessary.

To demonstrate successful transfer, we construct and use an entanglement witness (similar to *(29)*), which verifies entanglement if the sum of two visibilities in two mutually unbiased bases is above the classical bound of $\frac{\sqrt{2}+1}{2} \approx 1.21$ *(22,29)*. The data for the visibilities was taken in addition to the fringe measurements (apart from *l=+-300*) with longer integration: For the asymmetric OAM state *l=±10/±100* we achieved a value of 1.48(1). When both photons were transferred to *l =±100*, the witness was 1.55(1). Both values were calculated without any correction of the data and violate the classical limit by around 30 standard deviations, demonstrating the successful entanglement transfer. Because of the significantly smaller creation and detection efficiencies and therefore lower pair detection rate for *l =±300*, we corrected for accidental coincidence counts *(22)* resulting in a value of 1.6(3) for our entanglement witness. With a statistical significance of more than 80% we thus violate the bound for separable states with photons that each carry *l =±300* quanta of OAM. To further corroborate this successful entanglement creation, we transferred only one photon to *l=±300* and measured the other in the polarization bases. The measured witness was 1.628(4). Therefore, our results demonstrate that single photons can carry 300ℏ of OAM and that entanglement between two photons differing by 600 in quantum number can be achieved. Even in classical optics the highest value of OAM that had been created with an SLM was *l=200 (30)*.

Apart from the fundamental interest of entangling high quantum numbers, we also demonstrate the use of high OAM entanglement for remote sensing. For this we use the same method as before for creating (folded interferometric scheme including SLM) and analyzing (slit wheel method) high-OAM entangled states. When we transfer one photon to high OAM values and keep the other in its polarization state, the pair can be used to remotely measure an angular rotation with a precision, which is increased by the factor *l* compared to the situation when only polarization entangled photon pairs are used (Fig.4) *(22)*. This can lead to notable improvements for applications in the field of remote sensing, especially where low light intensities are required, like biological imaging experiments of light sensitive material. An analogous improvement can be achieved classically if diagonally or circularly polarized light enters our transfer setup. However, the important difference is that due to entanglement the measurements can be done remotely with the photons being space-like separated or even in unknown location at some later time.

Our approach could be generalized to higher dimensional entanglement for spatial modes, e.g. by starting with higher dimensional (hybrid) entanglement and a more complex interferometric scheme with potential benefits in applications like quantum cryptography, quantum computation and quantum metrology.

**Acknowledgements**

This work was supported by the ERC (Advanced Grant QIT4QAD, 227844), and the Austrian Science Fund FWF within the SFB F40 (FoQuS) and W1210-2 (CoQuS).

R.F. participated in the design and building of the experimental apparatus, collected and analyzed the data and wrote the manuscript. R.L., C.S. and S.R. participated in the design and building of the experiment and assisted on the experimental side. W.P., S.R. and M.K. assisted on the theoretical side. A.Z. initiated the work and supervised the experiment. All authors contributed to conceiving the experiment, discussing the results and contributing to the final text of the manuscript.

**Supplementary Materials**

**Materials and Methods**

In our experiment the polarization-entangled photon pairs were created using a 15mm long type-II nonlinear crystal (periodically poled potassium titanyl phosphate (ppKTP)) in a Sagnac-type configuration *(26,27)*. The crystal is pumped by a blue 405nm continuous-wave diode laser with up to 35mW of power. The down-converted pairs were filtered by a 3nm band-pass filter and coupled into single-mode fibers leading to approximately 1.3 million pairs per second. To ensure no polarization change between the source and the transfer setups, fiber polarization controllers were used.

The transfer setups were built with 50mm optics to be able to increase the beam waist such that even the outer region on the SLM, where more pixels per $2\pi$ phase shift are available, can be used. The SLM was not only used to transfer the photons from the Gauss mode to the Laguerre-Gauss mode, but also to optimize the spatial shape and therefore the overlap of the two created modes. This was accomplished by optimizing the circular shape and size with the help of cylindrical Fresnel lenses additionally displayed on the SLM.

After the transfer setup the spatial modes of the photons were adjusted to fit in size to the slit masks which had a diameter of approximately 25mm for 20 slits (*l=10*), 50mm for 200 slits (*l=100*) and approximately 150mm for 600 slits (*l=300*). The different sizes of the masks arise because of the low resolution of the laser cutter (approx. 100μm) that was used to cut the slits into the black paper while still achieving the same relative slit width (see discussion below for more details). The transmitted photons after the mask were focused to free-space single-photon detectors based on avalanche photo diodes with a circular active area of 500μm in diameter. The pair events were identified by a time-to-amplitude-converter with an effective coincidence window of 1.4ns.

## Discussion of the limiting effects

The $2\pi$ phase change in the outermost circumference of the displayed pattern corresponds to approximately 3000 pixels. This relates for example to only 10 pixels per $2\pi$ phase change for an OAM of $l=300$. In the inner area of the phase pattern there are even less pixels per $2\pi$ change available and Moiré patterns start to appear due to aliasing effects (Fig. 1D).

For entanglement in polarization a value of 1.9597(3) for the entanglement witness of equation (4) was measured (where the number in the brackets denote the statistical uncertainty assuming Poissionian fluctuations). The difference between the entanglement before the transfer to the very high OAM quanta and afterwards, can mainly be explained by the inaccurate conversion of the photons due to the aforementioned finite pixel sizes of the SLM display (see main text for more details). Additionally, an inevitable decrease in the measured correlations results from the measurement technique itself. Since the superposition structure, which follows a sin-function, never vanishes except in an infinitesimally small region, only an infinitesimally small slit width would lead to perfect fringes with 100% visibility. On the other hand, very small slit widths reduce the number of transmitted photons, so there is a trade-off between the theoretically measurable visibility of the fringes and sufficiently high count rates. The masks we used for $l=10$, $l=100$ and $l=300$ had a ratio of slit width to distance between two slits of approximately 1/7.1, 1/5.7 and 1/6.9 respectively. These ratios of the slit widths to the distance between two slits reduce the theoretical fringe visibility to around 96.8 (6) %, 95.0 (3) % and 96.6 (7) % respectively. By taking into account the imperfectness of the original polarization entangled photons (visibility 97.99 (3) %) the maximally achievable visibility for $l=10$, $l=100$ and $l=300$ would be of 94.8 (6) %, 93.1 (3) % and 94.7 (7) %. Misalignments of the transfer setup and masks additionally lower the measureable visibilities.

## Detailed calculation of the entanglement witness

In order to prove that the transfer of entanglement from polarization to OAM was successful, we calculate the entanglement witness *(29)* which consists of the sum of two visibilities in two mutually unbiased bases

$$\widehat{W} = \text{vis}_{\gamma 1} + \text{vis}_{\gamma 2} \ . \tag{1S}$$

The measurement technique with the slit mask can distinguish between any equally weighted superposition OAM states

$$|\chi\rangle = \frac{1}{\sqrt{2}} \left( |l\rangle + e^{i\varphi} |\text{-}l\rangle \right), \tag{2S}$$

where the phase between the OAM quanta $l$ and $-l$ is directly connected to the angle of the mask via the formula $\gamma = \frac{\varphi}{2l} \frac{360°}{2\pi}$. Therefore, each of the two visibilities of the witness operator can be rewritten in terms of two projections for different angular positions and hence the witness becomes

$$\widehat{W} = \hat{P}_{A\gamma 1, B\gamma 1} - \hat{P}_{A\gamma 1, B\gamma 1^{\perp}} + \hat{P}_{A\gamma 2, B\gamma 2} - \hat{P}_{A\gamma 2, B\gamma 2^{\perp}} \,, \tag{3S}$$

where A (B) stands for one of the two photons, $\gamma_1$ depicts the angular position of the first mutually unbiased basis. The angle $\gamma_2$ equals $\gamma_1 + \frac{45°}{l}$ and therefore stands for the second mutually unbiased basis. The $\perp$-sign illustrates the angular position of the respective orthogonal superposition e.g. $\gamma_{1\perp} = \gamma_1 + \frac{90°}{l}$ which is necessary to measure the visibility. To find the upper limit which is achievable with separable states we use the general pure separable 2 photon OAM state

$$|\psi\rangle = \left(a|l\rangle + be^{i\varphi_1}|-l\rangle\right) \otimes \left(c|l\rangle + de^{i\varphi_2}|-l\rangle\right) \,, \tag{4S}$$

with $a, b, c, d, \varphi_1, \varphi_2 \in \mathbb{R}$, $a^2 + b^2 = 1$, $c^2 + d^2 = 1$ and $l$ denotes the quanta of OAM. The straightforward calculation of the witness (3S) for the separable state (4S) leads to

$$\widehat{W} = cd\bigl(\cos\varphi_2\,(1 + 2ab\cos\varphi_1) + \sin\varphi_2\,(1 + 2ab\sin\varphi_1)\bigr) \,. \tag{5S}$$

Therefore, the maximal value of the witness $\widehat{W}$ for all separable states is

$$\widehat{W} = \frac{\sqrt{2}+1}{2} \approx 1.21 \tag{6S}$$

for $a=b=c=d=\frac{1}{\sqrt{2}}$ and $\varphi_1 = \varphi_2 = \frac{\pi}{4}$. If the sum of the visibilities is bigger than 1.21 the measured state is non-separable or in other words entangled.

**Data and additional measurements**

Correction of the data for accidental coincidences

A coincidence count is defined as two photons arriving at the two detectors within a defined coincidence window (in our experiment approximately 1.4ns). A combination of multiple down conversion events and intrinsic dark counts of the detectors within one timing window results in an 'accidental' coincidence rate. The expected accidentals can be approximated as the product of the timing window and the single counts at each detector. Due to low creation and detection efficiency for $l=\pm 300$ OAM entangled photons the achieved and presented results have to be corrected for accidentals. For all other measurements no correction is necessary.

Detailed table of performed measurements

For a complete description of the achieved measurement results, the measured visibilities for the states as described in the main text are attached in table S1. Additionally, the formula for the entanglement witness and the calculated value for each state are included.

entanglement witness:

$$\widehat{W} = vis_{\gamma 1} + vis_{\gamma 2} \begin{cases} \leq \frac{\sqrt{2}+1}{2} \approx 1.21 & separable \\ > \frac{\sqrt{2}+1}{2} \approx 1.21 & entangled \end{cases}$$

| state | $vis_{\gamma 1}$ | $vis_{\gamma 2}$ | $\widehat{W}$ |
|---|---|---|---|
| $\lvert 10\ -100\rangle + \lvert -10\ 100\rangle$ | 0.750(6) | 0.725(6) | 1.48(1) |
| $\lvert 100\ -100\rangle + \lvert -100\ 100\rangle$ | 0.772(5) | 0.776(5) | 1.55(1) |
| $\lvert H\ -300\rangle + \lvert V\ 300\rangle$ | 0.829(3) | 0.798(3) | 1.628(4) |
| $\lvert 300\ -300\rangle + \lvert -300\ 300\rangle$ (corrected data) | 0.9(2) | 0.7(2) | 1.6(3) |

**Table S1.** Measurements of the visibilities including all statistical uncertainties, which were calculated assuming Poissionian statistics. The witness W indicates entanglement for all states, although the statistical significance for l=±*300* with 1.33 standard deviations shows that the technical limit due the low creation and detection efficiency was reached (table row 4). To demonstrate more clearly that no fundamental limit was observed, the hybrid entangled state between polarization and l=±*300* was created and measured very precisely. A clear violation of more than 100 standard deviations was found.

Additional measurements

In order to test if the measured fringes at small rotation angles as seen in the main text continue to appear for bigger rotations, we measured the fringes in the coincidences in a rotation of 90 degree when both photons are transferred to $l=\pm 100$ (Fig. S2).

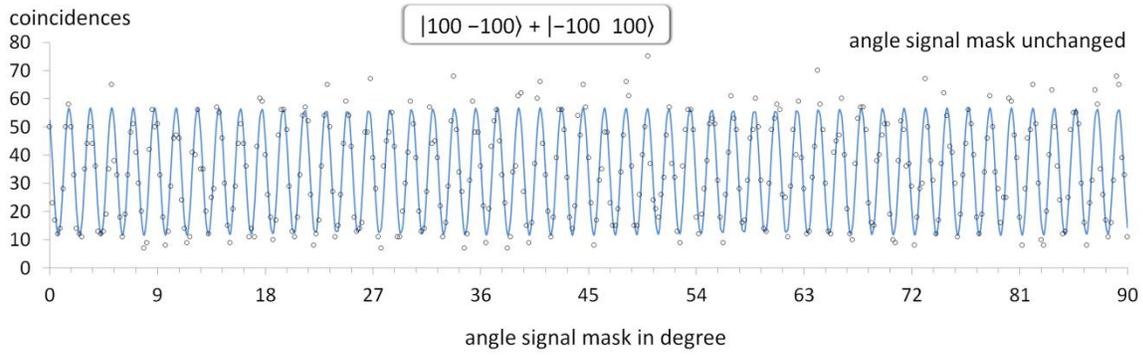

**Fig. S1.** Measurements of non-classical coincidence counts as a function of the rotation angle of one mask while the second mask is unchanged. The measured points show the expected 50 fringes for a rotation from 0° to only 90° for the rotated mask. Each data point corresponds to 30 seconds of measuring time and the line is the best fitted sin-function. Poissionian errors are not shown for the sake of clarity.

**Description of the OAM induced enhancement in angular precision**

Hybrid polarization-OAM-entangled states can be used to remotely measure an angular displacement with increased sensitivity. As mentioned in the main text, the physical orientation $\gamma$ of the mask the transferred photon passes, and the phase $\varphi$ of the superposition state which the mask selects, are connected via the equation $\gamma = \frac{360°}{2\pi} \frac{\varphi}{2l}$. Consequently, the OAM quantum number $l$ acts as an inverse scaling factor, where the angle of the rotated mask in the arm with the high OAM quanta can be measured with increased precision. For example, to keep the coincidence count rate constant (within the statistical fluctuation) during the rotation of the mask, the polarizer in the path of the unchanged photon would need to be rotated by angles which are larger by a factor $l$. In Fig. 4 in the main text, this enhanced precision is shown experimentally.